\documentclass[11pt]{article}

\usepackage{epsfig}
\usepackage{latexsym}
\usepackage{enumerate}
\usepackage{url}
\usepackage{float}
\usepackage[hypertexnames=false,hyperfootnotes=false]{hyperref}
\usepackage{texnansi}
\usepackage{color}
\usepackage{tikz}
\usepackage[margin=10pt,font=small,labelfont=bf]{caption}
\usepackage[subrefformat=parens,labelformat=parens]{subcaption}
\usepackage{afterpage}
\usepackage{enumitem}
\usepackage[boxed]{algorithm}
\usepackage{algpseudocode}
\usepackage[normalem]{ulem}
\usepackage{lmodern}
\usepackage{booktabs}
\usepackage{sectsty}
\usepackage{ifthen}
\usepackage{amsmath}
\usepackage{amsthm}
\usepackage{amssymb}
\usepackage{amsfonts}
\usepackage{thmtools}
\usepackage{thm-restate}
\usepackage{mathtools}
\usepackage{xspace}
\usepackage{titling}
\usepackage{natbib}
\usepackage{xfrac}
\usepackage{multirow}
\usepackage{bigdelim}
\usepackage{bm}
\usepackage{cleveref}
\usepackage[textsize=tiny]{todonotes}
\declaretheoremstyle[headfont=\sffamily\bfseries,bodyfont=\itshape]{thm-sf}

\crefname{assumption}{assumption}{assumptions}

\renewcommand{\thmcontinues}[1]{\hyperref[#1]{continued}}

\usetikzlibrary{arrows,patterns,plotmarks,pgfplots.groupplots}
\tikzstyle{every picture} += [>=stealth]
\tikzset{axis/.style={semithick, line join=miter}}
\allsectionsfont{\sffamily}
\makeatletter
\def\@seccntformat#1{\csname the#1\endcsname.\quad}
\makeatother

\floatname{algorithm}{\normalfont\sffamily\bfseries Algorithm}
\floatstyle{ruled}
\newcommand{\emailhref}[1]{\href{mailto:#1}{\tt #1}} 
\provideboolean{fastcompile}
\newcommand{\hidefastcompile}[1]{\ifthenelse{\boolean{fastcompile}}{}{#1}}
\usepackage{pgfplots}
\usepackage{pgfplotstable}
\usetikzlibrary{calc}
\usepackage{mathtools}
\definecolor{orange}{rgb}{0.85,0.33,0.13} 
\definecolor{green}{rgb}{0.13,0.85,0.33}
\definecolor{purple}{rgb}{0.33,0.13,0.85}
\definecolor{lime}{rgb}{0.65,0.85,0.13}
\definecolor{blue}{rgb}{0.13,0.65,0.85}
\pgfplotscreateplotcyclelist{tricolor}{%
  orange,every mark/.append style={fill=orange!80!black},mark=*\\%
  green,every mark/.append style={fill=green!80!black},mark=square*\\%
  purple,every mark/.append style={fill=purple!80!black},mark=otimes*\\%
  black,mark=star\\%
  orange,every mark/.append style={fill=orange!80!black},mark=diamond*\\%
  green,densely dashed,every mark/.append style={solid,fill=green!80!black},mark=*\\%
  purple,densely dashed,every mark/.append style={solid,fill=purple!80!black},mark=square*\\%
  black,densely dashed,every mark/.append style={solid,fill=gray},mark=otimes*\\%
  orange,densely dashed,mark=star,every mark/.append style=solid\\%
  green,densely dashed,every mark/.append style={solid,fill=green!80!black},mark=diamond*\\%
}
\pgfplotsset{colormap={tricolormap}{color=(orange) color=(green) color=(purple)},
  colormap={quadcolormap}{color=(orange) color=(lime) color=(blue) color=(purple)}}
\pgfplotstableset{%
  font=\small,
  every head row/.style={before row=\toprule[1pt], after row=\midrule},
  every last row/.style={after row=\bottomrule[1pt]}}
\pgfplotsset{compat=1.15}

\usepackage{setspace}
\usepackage[margin=1in]{geometry}

\provideboolean{submissionversion}
\setboolean{submissionversion}{true}
\provideboolean{blindversion}

\ifthenelse{\boolean{submissionversion}}{
  \renewcommand{\todo}[2][1]{}
  
  \newcommand{\deledit}[1]{}
}{
  
  \newcommand{\deledit}[1]{{\color{orange} \sout{#1}}}
}

\ifthenelse{\boolean{blindversion}}{
  \title{\bf\sffamily Price Action in Coupled Markets}
  \author{
    Althea Sterrett \\  Whetstone Research \\ email: \emailhref{althea@whetstone.cc} \and
    Austin Adams \\ Whetstone Research \\
    email: \emailhref{austin@whetstone.cc}
  }
  \date{}
}{
  \title{\bf\sffamily A Microstructure Analysis of Coupling in CFMMs\thanks{The authors wish to thank
      \href{https://x.com/js_horne}{Jacob Horne} and \href{https://x.com/bridgettfrey}{Bridgett Frey} for helpful comments.}}
  \author{
    Althea Sterrett \\  Whetstone Research \\ email: \emailhref{althea@whetstone.cc} \and
    Austin Adams \\ Whetstone Research \\
    email: \emailhref{austin@whetstone.cc}
  }
  \date{Initial version: October 7, 2025 \\ This version: October 7, 2025}
}

\begin{document}
\maketitle
\singlespacing

\begin{abstract}
  The programmable and composable nature of smart contract protocols has enabled the emergence of novel market structures and asset classes that are architecturally frictional to implement in traditional financial paradigms. This fluidity has produced an understudied class of market dynamics, particularly in coupled markets where one market serves as an oracle for the other. In such market structures, purchases or liquidations through the intermediate asset create coupled price action between the intermediate and final assets; leading to basket inflation or deflation when denominated in the riskless asset. This paper examines the microstructure of this inflationary dynamic given two constant function market makers (CFMMs) as the intermediate market structures; attempting to quantify their contributions to the former relative to familiar pool metrics such as price drift, trade size, and market depth. Further, a concrete case study is developed, where both markets are constant product markets. The intention is to shed light on the market design process within such coupled environments.
\end{abstract}

\section{Introduction}

Programmatic markets have created a new language for market creation. This is done by enabling creators and market designers to rapidly experiment with new and novel asset classes. One impact from this growth of programmable markets is the commodification of issuance. Asset issuance used to be an arduous challenge on all fronts, historically with fixed costs of inclusion into popular market destinations.

Additionally, programmatic markets have unlocked net-new designs that are not feasible with manual systems. For example, long-tail markets and what we refer to as "coupled markets". Coupled markets are a new primitive where the output to one market is used as the input to another - effectively coupling the prices of the assets together. This effect is easier to examine in markets where the pairing is an “oracle market”, where the price feed for the asset is solely determined by the AMM. Most markets trade in the most liquid venue - allowing an asset coupling effect by denoting the sale of an asset in another and placing them in an automated market maker.

In an automated market maker, the exchange rate can only be changed by trading against the pool. Because for (large enough) price deviations, there exists an arbitrage against the true value of this asset (generally assumed to be a centralized exchange) \cite{milionis2025automatedmarketmakingarbitrage}. This arbitrage enables a class of users who keep AMM prices inline with the "true value" of the asset by constantly arbitraging the market \cite{lehar2025decentralized}. 

In a central limit order book, the traders must both price the asset and provide liquidity for the market (via limit orders). By unbundling these two functions, new markets can be created by allowing sophisticated counterparties to only compete on their ability to price an asset - potentially at both low and high-frequencies. One simplifying assumption of AMM models is that there exists an external liquid venue which functions as the “true” market price \cite{milionis2024automatedmarketmakinglossversusrebalancing}. This assumption is likely true for the most well-known assets, but due to the growth of programmatic issuance it is being challenged for newer assets.

A key question emerges when the AMM is the liquid venue. This could emerge from the pool could be the only tradeable market for this asset or liquid enough to force asset coupling through carrying costs. We label this market as an "oracle market", as the “true” price of the asset is derived from this market, effectively making it an oracle. The oracle markets have unique properties, some of which we will explore today within asset coupling markets.

First, the delineation between noise and informed traders becomes meaningless. All traders functionally become informed, because there exists no market to arbitrage the price against. It is possible for some informed party to reset the market back (making the previous trade uninformed), but this is not without risk, as there exists no liquid venue to fully offset the costs. Constant fee AMMs treat every participant as informed (by charging every participant the same fee), making them well suited for this type of trading. 

Secondly, because of the lack of "noise trading", this market is almost exclusively being adversely selected. In traditional markets, this would leading to liquidity drying up, as traditional market makers require some amount of non-toxic flow to continue pricing assets. However, the asset issuer greatly benefits from the pricing information provided by the pool, and is thus greatly incentivized to provide long-term liquidity in the market. The value of pricing information is generally disregarded in most LPing models, because only the issuer is incentivized to pay this cost (and functionally does using market making deals).

We argue that oracle markets are much more likely to occur when combined with issuance protocols, such as \href{https://doppler.lol/whitepaper.pdf}{Doppler}, because liquidity is placed into the pool to abstract most of the complexities with market making. This makes the “canonical pool” (the pool where the issuance protocol houses most liquidity) an oracle market almost by design. It is possible as the token project grows that other venues may emerge (and we have seen this occur). This generally occurs because of high fees in the canonical pool, but issuance protocols have responded with dynamic fees to lower the spread.

Oracle markets may have more emergent properties as their place in the ecosystem evolves. 

In this paper, we explore asset coupling within oracle markets. By making the output currency of one oracle market as the input to another, strange properties arise, because the informed price update in the first oracle market effectively adds value to the second oracle market via utilizing that currency as their pair. 

This asset coupling strategy has been a popular way to drive value to arbitrary assets by utilizing \href{https://aada.ms/pdfs/dlb.pdf}{issuer monopoly}. By specifying the specific asset that value is traded in, the auction effectively becomes priced in the desired asset and has a coupling effect. Historically, asset issuers want to denote this in some widely used currency like the dollar or native tokens like ETH, as the ease of use in bidding in auctions far outweighs the benefit of an alternative currency. 

However, this problem is fixed due to guaranteed market linking via AMMs. With these links, arbitrary liquidity can flow into the pool because of a guaranteed market link between the issuance protocol and the market. By denoting the asset in the market, you create a programmable economic link between the two assets. The nature of the liquidity placement determines this link and can be adjusted to target specific outcomes.

\section{Analysis}

Suppose we have three different assets $x$, $y$ and $z$, and two constant function market makers, or CFMMs,  $\varphi_1,\varphi_2$ with trading pairs $y$/$x$ and $z$/$y$ respectively. If both CFMMs are the only $y$ and $z$-markets available, a positive coupling condition arises between the price of $y$ and the price of $z$ during purchase or liquidation events between assets $x$ and $z$. Any such trade will require routing through both intermediary markets, inherently dragging the price of $y$ along in the direction of the trade. This creates compound slippage and price action that inflates or deflates the received basket's value relative to the $x$ base.

To quantify coupling effects, we establish a baseline where the $y$/$x$ market has sufficient depth that $z$ market activity negligibly impacts $y$'s price. This approximates the scenario where $y$ serves as a stable numeraire for $z$ pricing - analogous to how ETH functions for small-cap token markets, where realistic trade volumes don't meaningfully move the ETH/USDC price. The intention of this baseline is to isolate the marginal effect of finite $y$/$x$ liquidity and the slippage it creates. For large-cap assets where this assumption breaks, the coupling effects we derive represent an upper bound on the inflationary dynamics.

The key point worth noting here is that if the consumption of $y$ and $z$ liquidity is correlated to changes in their respective market prices, then the potential for these inflationary or deflationary effects remains when coupled. This correlated consumption of liquidity could occur when underlying asset prices themselves are correlated, likely due to asset-specific effects. Hence, using the decoupled scenario as the analytic baseline allows for the direct isolation of the effect of coupled price action on the trader's portfolio. 

Quantifying the effects of coupling on the output basket, and how the underlying market structures contribute to this behavior, requires defining the following trader metrics:
\begin{itemize}
    \item[i.] The basket value discrepancy between coupled and decoupled scenarios on purchase and liquidation, both in the intermediate and output baskets.
    \item[ii.] The price drift transmission (how the price drift incurred in the initial swap market affects the drift incurred in latter markets), allowing us to see the exact coupling in price drifts in the $y$ and $z$ markets.
    \item[iii.] The marginal cost of additional swap liquidity as a function of the trade sizes, again on both purchase and liquidation.
    \item[iv.] The curvature of the compound $z$/$x$-market and what this implies about the compound market depth
\end{itemize}

These four metrics allow direct determination of the net inflation on the output value due to the coupling, how sensitive that inflation is to additional swap liquidity or price drift, and the dependencies these outcomes have on the depths of each swap market.

\section{Market Metrics}
\label{sec:mmetrics}

Consider a pool described by pair $(\mathbf{R},\varphi(x,y))$ where $\mathbf{R}=(x,y)$, $x$ is the base reserve, $y$ the quote, and $\varphi(\mathbf{R})=k$ is the invariant functions of the pool. The spot price, denominated in the market's base asset, follows the instantaneous marginal exchange rate of the pool reserves.
\begin{align*}
    P_x(\mathbf{R}) = \frac{\frac{\partial\varphi}{\partial x}}{\frac{\partial\varphi}{\partial y}}(\mathbf{R}) = -\frac{dy}{dx}(\mathbf{R}), && P_y(\mathbf{R})=-\frac{dx}{dy}(\mathbf{R}).
\end{align*}
Note that this only gives the reported price, not the actual cost of a swap. Consider the case where $\mathbf{R}=(x,y)$, with swap input amount $\Delta$ of $x$ and discount factor $\gamma$. The swap is computed by enforcing our invariant holds under a discounted token exchange~\cite{Angeris2021CFMM}. So the following equation must hold:
\begin{equation*}
        \varphi(x+\gamma\Delta,y-\Lambda) = k.
\end{equation*}
The actual price received on such a trade is simply $P_{eff}=\frac{\Delta}{\Lambda}$. This also shifts the pool price based on how we update the reserves during such a trade $\mathbf{R}\rightarrow\mathbf{R}'$. To get the marginal execution price for a trade of $\Delta$, we can just derive the above invariant:
\begin{align*}
    \frac{d}{d\Delta}\varphi(x + \gamma\Delta,y - \Lambda(\Delta)) &= \vec{\nabla}\varphi\cdot\left(\gamma \hat{x}-\frac{d\Lambda}{d\Delta}\hat{y}\right) = 0 \\
    \gamma\frac{\partial\varphi}{\partial x}&-\frac{\partial\varphi}{\partial y}\frac{d\Lambda}{d\Delta} = 0 \\
    \frac{d\Lambda}{d\Delta}(\Delta) &= \gamma P_x(\mathbf{R}')
\end{align*}
Notice that the marginal execution price is just the discounted price of the pool after the trade. As such, the relative price drift a market undergoes as a result of a trade can be expressed in terms of the marginal execution price for that trade.
\begin{align}
    \mu_x(\Delta) = 1-\frac{1}{\gamma P_x}\frac{d\Lambda}{d\Delta}(\Delta), && \mu_y(\Delta) = \frac{\gamma}{P_y}\left(\frac{d\Lambda}{d\Delta}(\Delta)\right)^{-1}-1.
    \label{eq:driftup}
\end{align}
It will also prove useful to derive their inverse relationships, as they can be used to study how these drift terms transmit through coupled markets.
\begin{align}
    \Delta(\mu_y)=P_y'^{-1}\left(\frac{P_y(\mu_y+1)}{\gamma}\right), && \Lambda(\mu_y)=\Lambda
    \left(P_y'^{-1}\left(\frac{P_y(\mu_y+1)}{\gamma}\right)\right).
    \label{eq:inverses}
\end{align}

Deriving the drift will give us the percent rate of change of the reported price as a function of swap input amount:
\begin{equation*}
    \frac{d\mu_y}{d\Delta}(\Delta) = -\frac{\gamma}{P_y}\frac{\frac{d^2\Lambda}{d\Delta^2}}{(\frac{d\Lambda}{d\Delta})^2}.
\end{equation*}
Now, the marginal swap depth, a measure of the rate of change of swap liquidity as a function of the price drift, is the reciprocal of this quantity.
\begin{equation}
    D_{\text{marg},y}(\mu_y) = -\frac{P_y}{\gamma}\frac{(\frac{d\Lambda}{d\Delta})^2}{\frac{d^2\Lambda}{d\Delta^2}}
    \label{eq:mdepth}.
\end{equation}
This quantity gives us a direct measure of the market's ability to absorb additional trade liquidity. So much like most measures of depth; the higher the value, the less shock that market will feel.

For the total depth of the market for a given price drift, simply integrate above marginal depth:
\begin{equation}
    D(\mu) = -\int_0^\mu\frac{P_y}{\gamma}\frac{(\frac{d\Lambda}{d\Delta})^2}{\frac{d^2\Lambda}{d\Delta^2}} d\mu_y
    \label{eq:depth}.
\end{equation}

This covers all the relevant CFMM metrics we utilize in our analysis.

\subsection{$z$-Purchase} 
\label{sec:zpurch}

To start, consider the difference of intermediate basket value between the coupled and decoupled scenarios:
\begin{align*}
    v_1&=P_y'\Lambda(\Delta)-\gamma\Delta \\ 
    &=\gamma\left(\frac{\Lambda(\Delta)-\Delta\frac{d\Lambda}{d\Delta}}{\frac{d\Lambda}{d\Delta}}\right)
\end{align*}
Given that CFMMs must remain concave, the price of $x$ after swap, $P_x'=\frac{d\Lambda}{d\Delta}/\gamma$, must obey the limit:
\begin{equation*}
    \lim_{\Delta\to\infty}\frac{d\Lambda}{d\Delta}/\gamma=0.
\end{equation*}
This implies that the intermediate value discrepancy itself must be unbounded, $v_1\to\infty$, with respect to trade size. Thus the concave nature of CFMMs (and realistically, most markets broadly) is the source of this basket value inflation relative to the decoupled scenario; the coupling only compounds this.

For the final $z$/$x$ trade output value, propagate the above amounts received through the $z$ market and compute the after trade prices for each scenario. Use the relation $P_y'=\gamma(\frac{d\Lambda}{d\Delta})^{-1}=P_y(\mu+1)$ along with the inverse functions \eqref{eq:inverses} to change the base coordinate to price drift in the $y$ market.
\begin{align*}
    v(\Delta) &= P_z'P_y'\Gamma(\Lambda(\Delta))-P_z'P_y\Gamma(\gamma_1\Delta/P_y)
\end{align*}
\begin{equation}
    v(\mu_y)=\gamma_2P_y\left[\frac{\Gamma(\Lambda(\mu_y))(\mu_y+1)}{\frac{d\Gamma}{d\Lambda}(\Lambda(\mu_y))} - \frac{\Gamma(\gamma_1\Delta(\mu_y)/P_y)}{\frac{d\Gamma}{d\Lambda}(\gamma_1\Delta(\mu_y)/P_y)}\right].
    \label{eq:valpurch}
\end{equation}

Notice that the only difference between the two compared cases, coupled as the initial term and decoupled as the latter, is the amount of tokens flowing into the $z$ market and the $y$ drift adjustment for the coupled scenario. Consider the effective trade price in this context, $P_{eff}=\frac{\Delta(\mu_y)}{\Lambda(\mu_y)}$. By concavity requirements, the effective price paid will be necessarily greater than or equal to the fee augmented spot price, $P_{eff}\geq\gamma_1P_y$, only equivalent in the case of constant sum markets. This would imply the inequality
\begin{equation*}
    {\frac{d\Gamma}{d\Lambda}(\gamma_1\Delta(\mu_y)/P_y)}>{\frac{d\Gamma}{d\Lambda}(\Lambda(\mu_y))}.
\end{equation*}
Which just means when denominated purely in $y$, the coupled basket comes out at a loss relative to the decoupled basket, which should be expected given the reduced $\Lambda$ input amount. This also gives us exactly the condition we want for determining when this coupling causes inflationary versus deflationary purchases relative to decoupled:
\begin{equation*}
    \mu_y\geq\frac{\Gamma(\gamma_1\Delta(\mu_y)/P_y)}{\Gamma(\Lambda(\mu_y))}\frac{\frac{d\Gamma}{d\Lambda}(\Lambda(\mu_y))}{\frac{d\Gamma}{d\Lambda}(\gamma_1\Delta(\mu_y)/P_y)}-1.
\end{equation*}

First note that because the quantity of $z$ received in the decoupled case is necessarily higher than the coupled case regardless of the $y$/$x$ CFMM at play, the right-hand side of this inequality is positive for all $\mu_y>0$. Meaning that it is theoretically possible for a CFMM to cause deflationary purchases when coupled in this setting, even with small purchases; the only requirement is that the price drift percentage is less than this right-hand side. Arranging this as an indicator function, for reasonable trade sizes, $\Delta\ll P_yy_1$ (or the equivalent price drift range), 
\begin{equation}
    l(\mu_y)=\mu_y+1-\frac{\Gamma(\gamma_1\Delta(\mu_y)/P_y)}{\Gamma(\Lambda(\mu_y))}\frac{\frac{d\Gamma}{d\Lambda}(\Lambda(\mu_y))}{\frac{d\Gamma}{d\Lambda}(\gamma_1\Delta(\mu_y)/P_y)}
\end{equation}
where $l>0$ implies inflation, $l<0$ implies deflation, and $l=0$ implies parity between coupled and decoupled behavior.

Next, for the price drift transmission, take the post-swap quantity from \eqref{eq:inverses} and propagate into the drift term \eqref{eq:driftup} for the $z$ market. 
\begin{equation}
    \mu_z(\mu_y)=\frac{\gamma_2}{P_z}\left(\frac{d\Gamma}{d\Lambda}\left(\Lambda_y\left(P_y'^{-1}\left(\frac{P_y(\mu_y+1)}{\gamma}\right)\right)\right)\right)^{-1}-1.
\end{equation}

This is a little opaque without a specific CFMM, but intuitively, this just computing the swap amount required to achieve a particular $\mu_y$, then propagating that amount through the computation of the price drift incurred in the second swap, $\mu_z$. Looking at the limiting behavior, as $\mu_y\to\infty$, the liquidity in the first pool is exhausted, leading to a bound on the input amount into the second swap. This means that $\mu_z(\mu_y)$ is always bounded based on the quantity of $y$ assets in the $y$ market, implying that transmission is bounded by this same amount. This is exactly the expected behavior for any potential $y$ market structure, not just CFMMs, due to the inherent finite liquidity constraint any realistic asset will encounter. Most of what is interesting about this function appears in its curvature, however, as this property would require a fixed CFMM function for proper analysis, it is left for the constant product market maker case study. 

Now, to study the marginal cost of purchase, consider a shared reserve state $(x, y_1, y_2,z)\in\mathbb{R}^4_+$. For a multi-asset trade $\mathbf{\Delta}$ with output $\mathbf{\Lambda}(\mathbf{\Delta})$, we can study the output received by looking at how this reserve state changes under a this trade's transition function, defined around the respective pool invariants:
\begin{equation*}
\mathbf{R}' = \mathbf{F}(\mathbf{R},\mathbf{\Delta})= (x+\Delta_x-\Lambda_x(\mathbf{\Delta}), y_1+\Delta_{y_1}-\Lambda_{y_1}(\mathbf{\Delta}), y_2+\Delta_{y_2}-\Lambda_{y_2}(\mathbf{\Delta}), z+\Delta_z-\Lambda_{z}(\mathbf{\Delta})),
\end{equation*}
with the definitions of $\Lambda_i(\mathbf{\Delta})$ fixed via solving for the swap output amount given the respective output reserve's invariant function. 

For the coupled scenario of interest, restrict $\Delta_{y_1}=\Delta_z=0$, and $\Delta_{y_2}'=\Lambda_{y_2}(\Delta_x)$. Since this transition function gives the reserves after swap, the marginal output of purchasing additional units is found by perturbing the swap input by $\delta\Delta$. So, consider the following expansion of $\mathbf{F}$. 
\begin{align*}
    \delta\mathbf{R}'&=\sum_{n=0}^\infty \frac{1}{n!}\frac{\partial^n\mathbf{F}}{\partial\Delta^n}(\delta\Delta)^n \\
    \delta\mathbf{R}'&= J_\mathbf{F}(\Delta)\delta\Delta+H_{\mathbf{F}}(\Delta)(\delta\Delta)^2 +\cdots \\
    \delta\mathbf{R}'&=\begin{bmatrix}
        1 \\ -\frac{d\Lambda}{d\Delta} \\ \frac{d\Lambda}{d\Delta} \\ -\frac{d\Gamma}{d\Lambda}\frac{d\Lambda}{d\Delta}
    \end{bmatrix}\delta\Delta + \frac{1}{2}\begin{bmatrix}
        0 \\ -\frac{d^2\Lambda}{d\Delta^2} \\ \frac{d^2\Lambda}{d\Delta^2} \\ -\frac{d^2\Gamma}{d\Lambda^2}(\frac{d\Lambda}{d\Delta})^2-\frac{d\Gamma}{d\Lambda}\frac{d^2\Lambda}{d\Delta^2}
    \end{bmatrix}(\delta\Delta)^2 + \cdots \\
    \delta\mathbf{R}'&=\begin{bmatrix}
        1 \\ -\frac{d\Lambda}{d\Delta} \\ \frac{d\Lambda}{d\Delta} \\ -\frac{d\Gamma}{d\Lambda}\frac{d\Lambda}{d\Delta}
    \end{bmatrix}\delta\Delta + \frac{1}{2}\begin{bmatrix}
        0 \\ \frac{\gamma_1}{P_y(\mu_y+1)^2D_{\text{marg},y}} \\ -\frac{\gamma_1}{P_y(\mu_y+1)^2D_{\text{marg},y}} \\ \frac{\gamma_2}{P_z(\mu_z(\mu_y)+1)^2D_{\text{marg},z}}\frac{d\Lambda}{d\Delta}^2+\frac{\gamma_1}{P_y(\mu_y+1)^2D_{\text{marg},y}}\frac{d\Gamma}{d\Lambda}
    \end{bmatrix}(\delta\Delta)^2 + \cdots
\end{align*}
Stop at second order for now, as it contains most of the information needed about curvature in the compound market. Because the only unit being purchased in the end in the $z$ asset, ignore everything but the fourth coordinate. In one line this reads as
\begin{equation*}
    \delta R_z'\approx -\frac{d\Gamma}{d\Lambda}\frac{d\Lambda}{d\Delta}\delta\Delta+\frac{1}{2}\left[\frac{\gamma_2}{P_z(\mu_z(\mu_y)+1)^2D_{\text{marg},z}}\frac{d\Lambda}{d\Delta}^2+\frac{\gamma_1}{P_y(\mu_y+1)^2D_{\text{marg},y}}\frac{d\Gamma}{d\Lambda}\right](\delta\Delta)^2.
\end{equation*}
Substituting in the price drift in the $y$-market, and changing the variables of the perturbation to be in terms of a change in drift, with $\delta\Delta=D_{\text{marg}, y}(\mu_y)\delta\mu_y$,
\begin{align*}
    \delta R_z'(\mu_y)\approx -&\frac{\gamma_1\gamma_2D_{\text{marg},y}}{P_zP_y(\mu_z(\mu_y)+1)(\mu_y+1)}\delta\mu_y \\
    &+\frac{1}{2}\left[\frac{\gamma_2D_{\text{marg},y}^2}{P_z(\mu_z(\mu_y)+1)^2P_y^2(\mu_y+1)^2D_{\text{marg},z}}+\frac{\gamma_1D_{\text{marg},y}}{P_y(\mu_y+1)^2P_z(\mu_z(\mu_y)+1)}\right]\left(\delta\mu_y\right)^2.
\end{align*}
Read another way, with $P'=P\cdot(\mu+1)$, the change in reserves in both $\Delta$ and $\mu_y$-spaces are the following:
\begin{equation}
    \delta R_z'\approx -\frac{\gamma_1\gamma_2}{P_z'P_y'}\delta\Delta+\frac{1}{2}\left[\frac{\gamma_2}{P_z'(\mu_z(\mu_y)+1)P_y'^2D_{\text{marg},z}}+\frac{\gamma_1}{P_z'P_y'(\mu_y+1)D_{\text{marg},y}}\right]\left(\delta\Delta\right)^2
    \label{eq:mcostpurchd},
\end{equation}
\begin{equation}
    \delta R_z'\approx -\frac{\gamma_1\gamma_2}{P_z'P_y'}D_{\text{marg},y}\delta\mu_y+\frac{1}{2}\left[\frac{\gamma_2D_{\text{marg},y}^2}{P_z'(\mu_z(\mu_y)+1)P_y'^2D_{\text{marg},z}}+\frac{\gamma_1D_{\text{marg},y}}{P_z'P_y'(\mu_y+1)}\right]\left(\delta\mu_y\right)^2.
    \label{eq:mcostpurchmu}
\end{equation}

Let's be more explicit about how this relates to the trader's output basket now:
\begin{equation*}
    \mathbf{O}(\Delta)=\mathbf{R}-\mathbf{R}'
\end{equation*}
with the marginal output as:
\begin{equation}
    o(\delta\Delta)=  -\delta\mathbf{R}'
\end{equation}
Now to get the marginal cost, all we need to do is flip this so:
\begin{equation}
    c(\delta\Delta) = -\frac{\delta\Delta}{\delta\mathbf{R}'}
\end{equation}

First thing worth noticing in \eqref{eq:mcostpurchmu}, there's a consistent factor of $(\mu_y+1)^{-1}$, implying that the marginal cost has a proportional relationship with the percent price adjustment in the $y$-market. This implies a cost structure that inflates relative to $x$ due to this $y$ dependence.

Now breaking down more intuitively what each term represents, \eqref{eq:mcostpurchd} and \eqref{eq:mcostpurchmu} can be summarized as:
\begin{equation*}
    \text{Marginal Output} = \text{Compounded Sensitivity } - \text{ Pool Curvature Penalties}
\end{equation*}
The linear term is just the product of marginal trade prices in each pool, and acts as a measure of the marginal price after trade on the compound $z$/$x$ market:
\begin{align*}
    a_1 &= \frac{d\Gamma}{d\Lambda}\frac{d\Lambda}{d\Delta}
\end{align*}
where
\begin{align*}
    \bigg|\frac{d\Lambda}{d\Delta}\bigg| = &\text{ marginal output sensitivity from the $z$-market}, \\ \bigg|\frac{d\Lambda}{d\Delta}\bigg|=&\text{ marginal output sensitivity from the $y$-market.}
\end{align*}
This would suggest the expected: intermediate CFMMs curves compound their slippage in the $z$/$x$ market.

For the second term of the expansion, let's define actual trade curvature for each pool as the second derivative of their swap methods:
\begin{align*}
    \kappa_y=-\frac{d^2\Lambda}{d\Delta^2}=\frac{\gamma_1}{P_y(\mu_y+1)^2D_{\text{marg},y}}, && \kappa_z=\frac{\gamma_2}{P_z(\mu_z+1)^2D_{\text{marg},z}}
\end{align*}
substituting these into our second-order term leaves us with:
\begin{equation}
    a_2 = -\frac{1}{2}\left(\frac{\kappa_y\gamma_2}{P_z'}+\frac{\kappa_z}{P_y'^2}\right)
    \label{eq:curvature}
\end{equation}
Notice what each term in $a_2$ represents. $\frac{\kappa_y\gamma_2}{P_z'}$ is just a product of the $y$-purchase curvature with the marginal execution price of purchasing $z$. In a sense, the contribution of how the $y$-market curvature contributes to the immediate curvature in the $z$-market. The latter term has the inverse interpretation, capture how the immediate marginal execution price in the $y$-market propagates into the purchase curvature on the $z$-market, where the squared factor arises due to the coupling the purchase has on the price of $y$. Thus, $-2a_2$, which is exactly the compound $z$/$x$ market's purchase curvature, is a coupled sum of the intermediate markets purchase curvatures. 

More practically speaking, \eqref{eq:curvature} shows that poor depth in \emph{either} pool creates outsized slippage in the compound market. From the perspective of a market designer, choosing where to concentrate liquidity isn't about just optimizing each pool's distribution in isolation - you need to balance depth across both pools.

\subsection{$z$-Liquidation}
\label{sec:zliq}

Consider now a $z$ liquidation event. The trader holds a basket of $\Gamma$ $z$ tokens and wishes to liquidate all of it for $x$ assets. Note that the swap functions will be flipped for this event, $\Lambda(\Delta)\to\Delta(\Lambda)$. 

First, use the relation $P_z=P_{z/y}=\frac{1}{P_{y/z}}$ to redefine the price drift in terms of familiar $x$ denominations:
\begin{align*}
    \mu_z=\mu_{y/z}(\Gamma)=1 - \frac{1}{\gamma_2P_z}\frac{d\Lambda}{d\Gamma}(\Gamma), &&
    \mu_{y}(\Gamma)=\gamma_2P_z\left(\frac{d\Lambda}{d\Gamma}(\Gamma)\right)^{-1}-1
\end{align*}
Similarly, on the $y$ market,
\begin{align*}
    \mu_{y_1}=\mu_{x/y}(\Lambda)=1-\frac{1}{\gamma_1P_y}\frac{d\Delta}{d\Lambda}(\Lambda), &&  \mu_x(\Lambda)= \gamma_1P_y\left(\frac{d\Delta}{d\Lambda}(\Lambda)\right)^{-1}-1
\end{align*}
To start, again consider the difference in basket value between coupled and baseline scenarios on just the $y$ sale alone:
\begin{align*}
    v_2=\Delta(\Lambda)-\gamma_1P_y\Lambda
\end{align*}
There is nothing special really happening here. Because $P_{y,eff}<\gamma_1P_y$ where $\Delta(\Lambda)=P_{y,eff}\Lambda$ its clear that $v_2<0$ for all $\Lambda$, simply implying the sale of tokens on a CFMM undergoes slippage. Now propagate the swap output from the sale of $\Gamma$ on the $z$/$y$ market through $v_2$ for the net value discrepancy:  
\begin{equation}
    v(\Gamma)=\Delta(\Lambda(\Gamma))-\gamma_1P_y\Lambda(\Gamma)
    \label{eq:vliq}
\end{equation}
Since this is fundamentally the same equation as above relative to the amount of $y$ assets being sold, the same deflationary principle persists. The main distinction will appear again in the compounded curvature of this function relative to $\Gamma$ as opposed to $v_2$ with $\Lambda$. It should also be noted the different form this takes relative to these metrics for the purchase event is simply from the choice of denomination being the target asset rather than the initial.

Now, using the same state space model as the purchase event, define a liquidation transition function by swapping in the opposite direction.
\begin{equation*}
    \mathbf{F}_l(\mathbf{R},\Gamma)= (R_{x}-\Delta(\Lambda(\Gamma)), R_{y_1}+\Lambda(\Gamma), R_{y_2}-\Lambda(\Gamma), R_{z}+\Gamma)
\end{equation*}
So, with a similar perturbative expansion:
\begin{align*}
    \delta\mathbf{R}'&=\sum_{n=0}^\infty \frac{1}{n!}\frac{\partial^n\mathbf{F}_l}{\partial\Gamma^n}(\delta\Gamma)^n \\
    \delta\mathbf{R}'&= J_{\mathbf{F}_l}(\Gamma)\delta\Gamma+H_{\mathbf{F}_l}(\Gamma)(\delta\Gamma)^2 +\cdots \\
    &=\begin{bmatrix}
        -\frac{d\Delta}{d\Lambda}\frac{d\Lambda}{d\Gamma} \\ \frac{d\Lambda}{d\Gamma} \\ - \frac{d\Lambda}{d\Gamma} \\ 1
    \end{bmatrix}\delta\Gamma + \frac{1}{2}\begin{bmatrix}
        -\frac{d^2\Delta}{d\Lambda^2}\left(\frac{d\Lambda}{d\Gamma}\right)^2-\frac{d\Delta}{d\Lambda}\frac{d^2\Lambda}{d\Gamma^2} \\ \frac{d^2\Lambda}{d\Gamma^2} \\ -\frac{d^2\Lambda}{d\Gamma^2} \\ 0
    \end{bmatrix}(\delta\Gamma)^2 + \cdots
\end{align*}
Isolating just for the $x$ reserve coordinate, and substituting in the curvature terms for each sale:
\begin{equation}
    \delta R' \approx -\frac{d\Delta}{d\Lambda}\frac{d\Lambda}{d\Gamma}\delta\Gamma+\frac{1}{2}\left[\frac{1}{\gamma_1P_yD_{\text{marg},x}}\left(\frac{d\Delta}{d\Lambda}\frac{d\Lambda}{d\Gamma}\right)^2+\frac{1}{\gamma_2P_zD_{\text{marg},y}}\frac{d\Delta}{d\Lambda}\left(\frac{d\Lambda}{d\Gamma}\right)^2\right](\delta\Gamma)^2.
\end{equation}
In terms of the relative marginal depths, with the change of variables $\delta\Gamma=D_{\text{marg},y}\delta\mu_y$ this reads as
\begin{equation}
    \delta R' \approx -\frac{\gamma_1\gamma_2P_yP_zD_{\text{marg},y}}{(\mu_y+1)(\mu_x+1)}\delta\mu_y+\frac{1}{2}\left[\frac{\gamma_1\gamma_2^2P_yP_z^2D_{\text{marg},y}^2}{(\mu_y+1)^2(\mu_x+1)^2D_{\text{marg},x}}+\frac{\gamma_1\gamma_2P_yP_z}{(\mu_y+1)^2(\mu_x+1)}D_{\text{marg},y}\right](\delta\mu_y)^2.
    \label{eq:mcostliqmu}
\end{equation}
Where the marginal cost function is again $c(\delta\Gamma)=-\frac{\delta R'}{\delta\Gamma}$. This is essentially the same result as the purchase scenario, with the only difference being direction (hence the reciprocal price appearances). Looking at the liquidation curvature terms for each pool now:
\begin{align*}
    \kappa_y = \frac{\gamma_2P_z}{(\mu_y+1)^2D_{\text{marg},y_1}}, && \kappa_x=\frac{\gamma_1P_y}{(\mu_x+1)^2D_{\text{marg},x}} 
\end{align*}
with the compound liquidation curvature being $-2b_2$ for
\begin{equation*}
    b_2 = -\left(\frac{\gamma_2^2P_z^2\kappa_x}{(\mu_y+1)^2}+\frac{\kappa_y\gamma_1P_y}{(\mu_x+1)}\right).
\end{equation*}

\section{Constant Product Market Maker Case Study}
\label{sec:univ2}

Constant product market makers (CPMM), such as Uniswap v2, is a protocol that facilitates spot trading markets. The constant product market maker, defined by the invariant
\begin{equation*}
    \varphi(x,y)=xy,
\end{equation*}
is one of the most well studied CFMMs to date. It boasts many nice properties for a spot market, such as readily available marginal trading liquidity at any price, for either asset in the pool, or extremely implementation-friendly core trading functions. Most importantly for this analysis, it provides a tractable setting to develop an intuition that readily extends to the general case, for how many CFMMs will behave.

Before the coupled analysis, start by computing the core pool metrics defined in (1) for a CPMM. The price functions are simply the ratio of reserves,
\begin{align*}
    P_x=\frac{\varphi_x}{\varphi_y}=\frac{y}{x}, && P_y = \frac{x}{y}.
\end{align*}
To compute the swap function, hold $\varphi(x,y)$ invariant under the trade and isolate for $\Lambda$:
\begin{align*}
    (x+\gamma\Delta)(y-\Lambda)=k \\
    y-\Lambda = \frac{xy}{x+\gamma\Delta} \\
    \Lambda(\Delta) =\frac{y\gamma\Delta}{x+\gamma\Delta}.
\end{align*}
Deriving this functions, and substituting into \eqref{eq:driftup}, yields  the price drift of both asset prices due to the trade:
\begin{align*}
    \mu_y=\frac{(x+\gamma\Delta)^2}{y^2}-1, && \mu_x = 1-\frac{x^2}{(x+\gamma\Delta)^2},
\end{align*}
with inverse function
\begin{equation*}
    \Delta(\mu_y) = \frac{y}{\gamma}\sqrt{\mu_y+1}-\frac{x}{\gamma}.
\end{equation*}
Deriving the drift more time and taking the reciprocal recovers the marginal trade depth \eqref{eq:mdepth},
\begin{equation*}
    D_{\text{marg},y}=\frac{2\gamma(x+\gamma\Delta)}{y^2}
\end{equation*}

This concludes the single pool CPMM metrics required to derive the relevant coupling behavior. Note for both purchase and liquidation events, all visualizations will be taken with respect to the same parameterization of the coupled system. So consider an auction house-like liquidity structure with the reserves $x=10^7$, $y_1=4.5\times10^8$, $y_2=7.2\times10^7$, and $z=10^9$, where both pool fees are set to $3\%$. While these fees are typically way too high for established markets, this structure mirrors what is seen in a lot of emerging markets, where the $y$-market rightfully holds the majority share of the $y$ liquidity relative to the $z$ market and the effect of coupling is significant.

\subsection{$z$-Purchase}
\label{sec:univ2purch}

For the $z$ purchase event on a coupled CPMM-CPMM market, start by computing the quantity of $z$ the trader receives,
\begin{equation*}
    z\text{ received}=\Gamma(\Delta)=\frac{y_1z\gamma_2\gamma_1\Delta}{xy_2+(y_2+\gamma_2y_1)\gamma_1\Delta}.
\end{equation*}
In the decoupled scenario, use the spot price of the initial market to fix the entry swap quantity in the $z$ market, so $\Gamma(\frac{y_1}{x}\gamma_1\Delta)$. 

Substituting the derived pool metrics in \eqref{eq:valpurch}, the basket value discrepancy as a function of price drift in the $y$ market reduces to
\begin{equation}
    v(\mu_y) = 
    \frac{\gamma_2x}{y_1y_2}\left(\frac{y_1}{x}\sqrt{\mu_y+1}-1\right)\left(\sqrt{\mu_y+1}(xy_2+xy_1\gamma_2 - \gamma_2y_1^3/x)+\gamma_2(y_1^2-x^2)-y_1y_2\right).
\end{equation}
Notice this is a quadratic function with respect to $u=\sqrt{\mu_y+1}$, with $v(0)=0$, $v'(0)=0$, and $v''(u)>0$ for $\forall u>0$; implying the $z$ purchase event yields always a basket with inflated value.

\begin{figure}[H]
    \centering
    \includegraphics[width=0.55\linewidth]{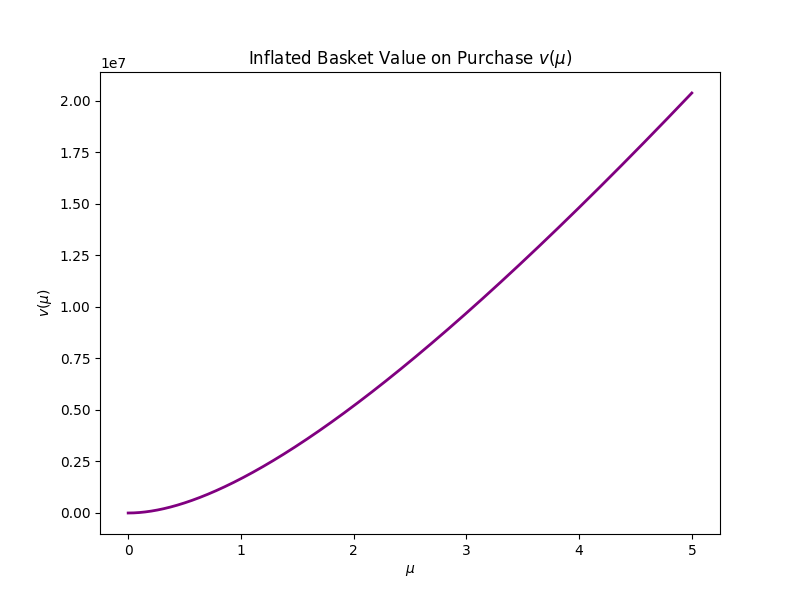}
    \caption{The basket value difference between the coupled and decoupled market scenarios on a $z$-purchase event.}
    \label{fig:pval}
\end{figure}

On a per-trade level, each pool experiences some amount of portfolio loss as the trade pushes the prices higher, often referred to as impermanent loss in this context. This loss is path-independent in the price process, implying that any change in valuation dynamic between trader and pool can be reversed (up to fee accounting). The liquidity provider's portfolio loss is exactly the source the above convex basket inflation, just sourced from two pools as opposed to one. 
 
\begin{figure}[H]
    \centering
    \includegraphics[width=0.55\linewidth]{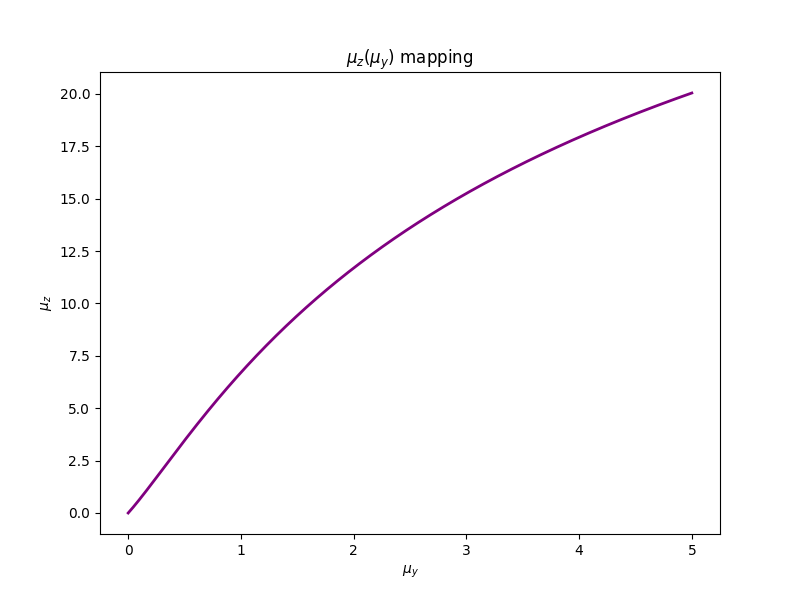}
    \caption{The transmission of $\mu_y$ into the price drift in the $z$-market, $\mu_z(\mu_y)$.}
    \label{fig:passthrough}
\end{figure}

Now turning to the price drift transmission,
\begin{align}
    \mu_z(\mu_y)&=\frac{(y_{2}x\sqrt{\mu_y+1}+\gamma_{2}y_{1}x(\sqrt{\mu_y+1}-1))^{2}}{x^2(\mu_y+1)y_{2}^{2}}-1.
    \label{eq:muzmuy}
\end{align}
For the provided reserve and fee values set at the start of the section,  \ref{fig:passthrough} displays this transmission function.

Notice the rate of change of this function is more strongly positive for small swap amounts, gradually slowing until the price coupling flattens out once most of the $y$ tokens have been purchased in the initial market. Whether this transmission is reflexive, however, where the drift $\mu_z$ denominated relative to $y$ outpaces $\mu_y$ depends entirely on how the $y$ liquidity is distributed among the intermediate pools.

Next, look at the marginal cost function in $\mu_y$-space. Explicit closed-form expressions for the CPMM marginal cost functions are available in the supplementary materials, though practitioners will find equation \eqref{eq:mcostpurchmu} and the derived swap, marginal depth, and price drift functions more tractable for numerical evaluation.
 
A topographic plot of $o(\mu_y,\delta\mu_y)$ is for the purchase event is seen in \ref{fig:mpurch} below.

\begin{figure}[H]
    \centering
    \includegraphics[width=0.55\linewidth]{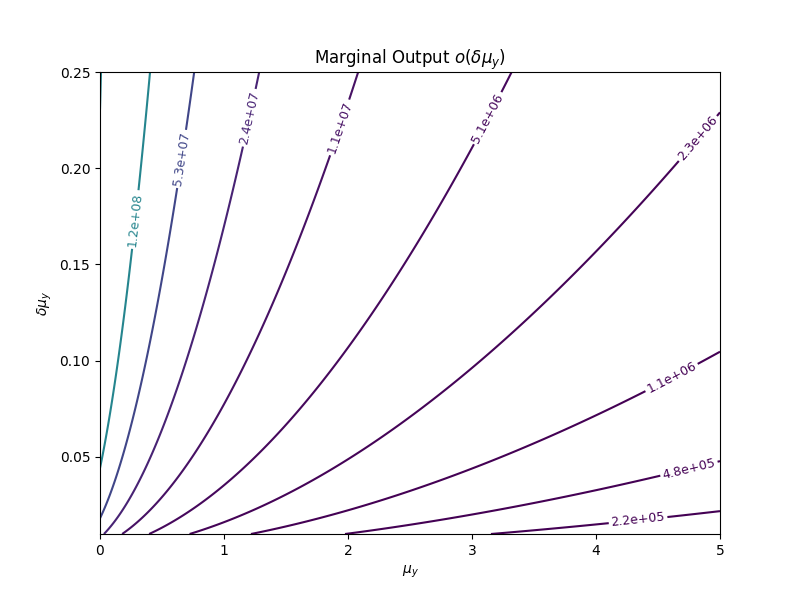}
    \caption{The marginal output amount, $o(\delta\mu_y)$, for $\delta\mu_y$ additional price drift in the $y$-market.}
    \label{fig:mpurch}
\end{figure}

As convexity of the CFMM invariant would suggest, the marginal output amount received is highest for small trade sizes diminishing with larger and larger size. As well, from what can be observed in the gradient, the rate that this marginal output decreases by also decreases with trade size. This is very much what we would expect from any CFMM that doesn't specifically fix these values (such as a constant sum curve would), implying that the compounded market behaves much like a CFMM in its own right when isolating the markets for just $z$/$x$ trades. This is also a supported conclusion in the general case derived in \ref{sec:zpurch} considering the compound market's path-independence, along with all the convexity requirements being met in the liquidity provider's portfolio and trader's basket value functions. In reality, however, so long as the $y$ market has it's own structure and external liquidity markets, the macroeconomic behavior of the compound $z$/$x$ market will never simplify to \emph{just} a CFMM. Instead, it may be observed that liquidity will percolate in and out of the market through the $y$ asset. 

\subsection{$z$-Liquidation}
\label{sec:univ2purch}
Next, the $z$-Liquidation scenario. First we should note we are flipping the direction of everything. So for swap functions:
\begin{align*}
    \Lambda(\Gamma)=y_2-\frac{zy_2}{z+\gamma_2\Gamma}=\frac{y_2\gamma_2\Gamma}{z+\gamma_2\Gamma} && \Delta(\Lambda)=\frac{x\gamma_1\Lambda}{y_1+\gamma_1\Lambda}
\end{align*}
Which leaves the final $x$ received to be:
\begin{equation*}
    x\text{ received}=\Delta(\Lambda(\Gamma)) = \frac{x\gamma_1y_2\gamma_2\Gamma}{y_1({z+\gamma_2\Gamma})+\gamma_1y_2\gamma_2\Gamma}
\end{equation*}

Start by formalizing the net value discrepancy, relative to price drift in $y$ on the $z$/$y$ market, $\mu_y$,
\begin{equation}
    v(\mu_y) = -\frac{x(\gamma_1y_2^2\sqrt{\mu_y+1}-\gamma_1y_2z)^2}{y_1(z+\gamma_2)(y_1y_2\sqrt{\mu_y+1}+\gamma_1y_2^2\sqrt{\mu_y+1}-\gamma_1y_2z)}
\end{equation}
Notice that for $\mu_y>0$, $v(\mu_y)<0$ implying a deflation in the basket value relative to the decoupled scenario, just as observed in the general case with \eqref{eq:vliq}. For the provided reserves and fee parameters set at the beginning of the case study, this function can be seen in figure \ref{fig:vliq} below.

\begin{figure}[H]
    \centering
    \includegraphics[width=0.55\linewidth]{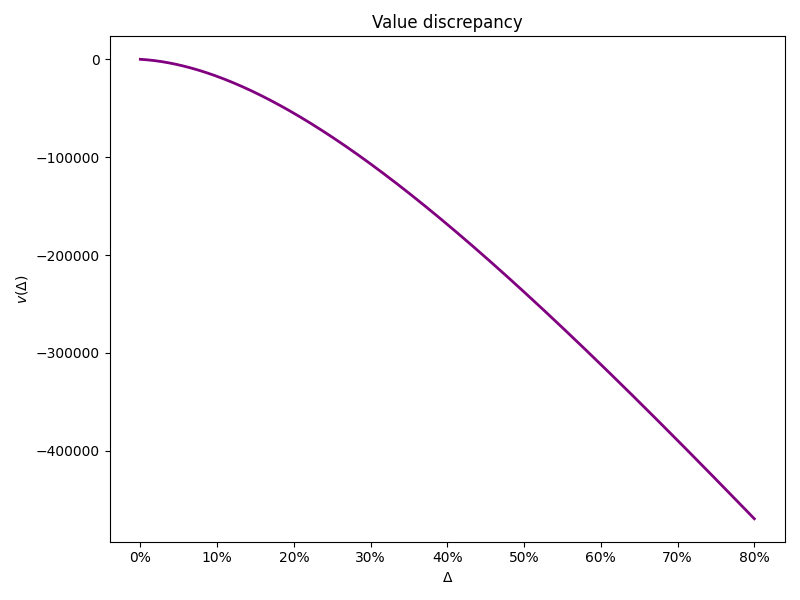}
    \caption{The basket value difference between the coupled and decoupled market scenarios on a $z$-liquidation event.}
    \label{fig:vliq}
\end{figure}

Recall where this value deflation is going as well. In the decoupled scenario, it's assumed the price of $y$ cannot be impacted; whereas in the coupled scenario, the second trade occurs on a regular CFMM. Both scenarios have the same trade outcome on the $z$ market, meaning the only place this deflation is going to is slippage on the $y$/$x$ market, or its liquidity providers to be more precise. This is why the effect is less compounded (by a factor of roughly $\sqrt{\mu_y+1}$) on liquidation rather than in purchases, given the fixed $x$ denomination. We highlight that this does not necessarily mean that there is an arbitrage opportunity here. Both markets are still robust - implying their compound is as well. In both scenarios, if one were to atomically perform the $z$ purchase and liquidation events to and from the $x$ asset, their net basket value would be roughly the same as their starting point, minus fees paid. The asymmetry in this value metric between both directions comes mainly from a difference in \emph{what} it's measuring rather than any inherent market asymmetry itself. 

Next, the price drift transmission function.
\begin{equation*}
    \mu_x(\mu_y) = \frac{(y_1z\sqrt{\mu_y+1}+\gamma_1y_2z\sqrt{\mu_y+1}-y_2z)^2}{y_1^2z^2(\mu_y+1)}-1
\end{equation*}
This is the exact mirror of \eqref{eq:muzmuy} in the opposite direction; same behavior and scaling with respect to the initial market's price drift. This is more of an artifact of the symmetry of constant product curves, leading to very symmetrical compound behavior as well in either direction (particularly when unnormalized by a particular denomination). 

Lastly, the marginal cost functional. Like before, in practice it will make more sense to work with the above derived quantities and \eqref{eq:mcostliqmu} directly. A topographic plot of this \eqref{eq:mcostliqmu} for the coupled CPMM case study, using the reserves and fees set at the beginning of the section, can be seen in figure \ref{fig:mliq} below.

\begin{figure}[H]
    \centering
    \includegraphics[width=0.55\linewidth]{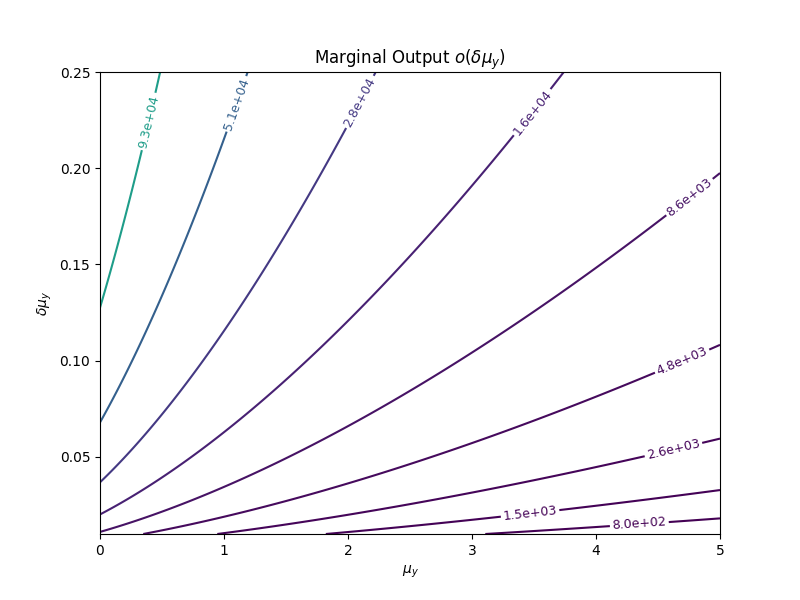}
    \caption{The marginal output amount, $o_{\text{liq}}(\delta\mu_y)$, for $\delta\mu_y$ additional price drift on the $z$-market.}
    \label{fig:mliq}
\end{figure}

Similar results as what we see in the purchase scenario, marginal output received is highest for small swap sizes with a gradual degradation in the marginal amount as the pool liquidity gets used. You'll notice that the gradient is much more gradual in this direction relative to the purchase scenario. This is nothing more than an artifact of the liquidity imbalances in the pools. Recalling our set up, the $y$/$x$ market holds most of the $y$ liquidity, and the $x$ token liquidity is in much lower \emph{quantity} than the $z$ liquidity in its respective market, which in this scenario manifests in a more gradual degradation in the liquidation direction.

While the liquidation metrics appear to algebraically mirror the purchase metrics, the incidence of value transfer is reversed. In purchases, convexity favors the trader through basket inflation, whereas in liquidations the same convexity shifts value back toward LPs, manifesting as systematic basket deflation. As well, this isn't so much a property of coupled markets as it is one of CFMMs themselves, only compounding in the coupling scenario to create a naturally leveraged derivative of this dynamic. Similarly, although price drift transmission follows the same structural form, the liquidity bottlenecks differ: deeper purchasable $y$ liquidity implies higher drift transmission, but whereas deeper $y$ liquidity in the final swap market implies a smoother gradient in the marginal cost and output amounts.

\section{Macro Implications}

Now that we have established the foundational equations, we will now briefly explore the direct implications of the market structure.

First, consider what happens when the $y$/$x$ market is the oracle market for the $y$ asset. There is no restriction on the amount of markets that use that $y$ base asset. However, the oracle market is main route between $y$ and it's derivatives with the numeraire. This means there is likely a large demand from non-toxic flow that will move through this route, as the price is correct more often than other markets \cite{adams2025amammauctionmanagedautomatedmarket}. As mentioned near the end of \ref{sec:univ2purch}, if $y$ is not a fully internalized asset, then there will be some form of liquidity percolation from both $x$ and $z$ out of their respectively paired $y$ reserves, especially as the number of $y$ denominated markets expand. The optimal design for a $y$-asset would mimic more of a currency than an asset.  Both of these will push their local market prices of $y$ upwards, for use by the trader somewhere else they may find utility for $y$. In the context where the $y$/$x$ market is an oracle market, this percolation can only flow upward rather than grounding back to the numeraire, as a form of rebalancing effects.

This implies an internalization of the natural inflation introduced in the coupled market scenario, as no arbitrage can realistically be established to push the price of $y$ in opposition of any traders through the oracle market onto the derivative markets. The base oracle market only deflates all traders baskets through the trickle of traders aiming to capitalize on their profits, or from some macro pressure on traders to no longer remain within the $y$ derivative ecosystem as a whole. This internalization is why, in practice, this cannot be treated as simply inflation that immediately gets deflated out of the system by arbitrage; flows stay internal unless pressured to reverse. While a trader's portfolio consists of $y$ derivative market assets, the portfolio will move with leveraged price action relative to the drift in the $y$ asset price, by a factor of $(\mu_y+1)$. It's more the oracle market dependence naturally introduces leverage for the whole $y$ derivative ecosystem, which is not dissimilar to other currencies.

It is possible that a market participant, such as a trader or asset issuer, could see the inflation in the $z$-market as a short-run opportunity to reset the price and capture additional value. This functions similarly to back-running arbitrage in traditional CFMM markets. However, as previously stated, this is not free as it requires users to take on significant price risk as the market exists in a fully liquid state of no-arbitrage (there is no-where to offload the risk). Additionally, because asset issuers are likely coupling assets with some type of informed price relationship, the belief in the correlation undoing itself would require some informed decision on the coupling changing or price movements. This functions similarly to outsourcing price risk via underwriting the loss with the value in the pool.

This coupling is not without downsides. It is more likely that internal leverage theoretically builds up in the system and then could potentially flushed swiftly during periods of financial malaise. This has been the focus of many financial papers specifically on financial crises (see Geanakoplos (2010) \cite{geanakoplos2010leverage}). Leverage is rapidly flushed out of a system, because asset holders become forced sellers into illiquidity, resulting in a large amount of value loss. 

However, we now stop to say that Geanakoplos is highlighting leverage in the 2008 financial crisis. Leverage in traditional financial institutions, such as GSIBs, grew to such a point that it was flushed. The argument here is rather that this risk being internalized is better due to less information asymmetry, more collectivization, and better internalization to believers in the system.

We argue that the key difference between the implied leverage in this system vs. others is how it is managed and by whom. In the coupled markets model, the markets themselves manage the leverage and the leverage is essentially programmatic. There is also less information asymmetry because all participants know the leverage in the system. Additionally, collectivization could enable more value demanders, perhaps many of which may not be able to access leverage. We argue that one of the large issues with leverage in traditional financial systems is that it is hidden. Every user is incentivized to leverage up as much as possible while also hiding it from other participants. This creates a similar effect to the tragedy of the commons where everyone is likely better off from not using leverage, but individual payoffs create a great incentive to deviate from the collective good strategy.

However, the cost of a leverage collapse is direct losses to users and should be avoided if possible. We argue that mechanics such as guaranteed liquidity, PID controllers, and slowing down the price evolution could programmatically limit the catastrophic effects of a forced deleverage event. On the other hand, this could increase the amount of moral hazard in the system \cite{cieslak2021economics}.

Because the leverage is programmatic and a direct result of the chosen curvature, it could be adapted over time in response to market events. We envision that this will be how issuance protocols and market participants adapt to this new technology. 


\subsection{Coupled Content Market Implications}

While the market described in this note applies to a lot of situations, one especially prominent case is in content markets such as with \href{https://zora.co/}{Zora Protocol}, the largest user of Doppler as an issuance protocol. Creators themselves have their own assets, which in this case is $y$, where their primary market is denominated in the \$zora token, our $x$. Further, creators make content that has its own fungible asset, the $z$ asset here, traded against the creator's asset. This is exactly the coupling we modeled above for the purchases and sales of the content asset from a user's basket of \$zora tokens. 

This coupling effect is used in content markets to quite literally couple the price paths of the assets together, creating a price relationship between two assets. This allows the creator's coin to function as partly a market of a basket of derivatives of their underlying content, coupling the asset prices together. Additionally, the content itself gains an economic relationship with the creator, allowing for content to potentially gain in value due to the derivative asset it functionally holds from the creator. 

We argue that this is a valuable relationship as it enables unique long-term pricing relationships. Additionally, this effect is only possible because of the implicit price correlation. With no price correlation, there $z$-market would be compressed to extract the added leverage. In reality, we argue that this coupling effect likely does impact larger assets, as there exists no infinite depth market. Token issuance has effectively locked up billions of dollars worth of assets to date. While it is impossible to say for certain, it has likely had enormous compounding effects on assets such as ETH and SOL, which are widely used as the liquid asset for token projects.

This relationship is programmatic and adjustable, allowing more or less price impact to be accrued as desired. Importantly, this price impact is not equal in both directions as the implied leverage effect is based on the depth of the $y-$ and $z-$markets. The deeper the market, the less effect on the asset's price. This phenomenon of asset coupling, as demonstrated by empirical results from Zora's content markets, testifies to the price coupling of these paired assets and the programmability enabled by issuance protocols like Doppler.

\subsection{Future Directions}

The above framework can be extended in a few ways to study the behavior of these markets further. The above only models the microstructure behavior of a single atomic swap sequence. Analyzing price discovery behavior more quantitatively will require a time series in the state space we can model with $\mathbf{F}$, $\mathbf{F}_l$, and the single pool swap transition functions. Further, it may merit a simple multi-agent simulation of this discovery process if the stochastic processes that arises becomes too burdensome to approach analytically.

Beyond this another simple extension, as seen by both the unnormalized marginal cost and value discrepancy functionals in the CPMM case study, we can directly observe the exact dependencies each variable has on each term as well as the final outcome. This suggest that, given some exotic CFMM with parameterizations other than just the fee, so:
\begin{align*}
    \varphi(\mathbf{R}, \alpha_0, \alpha_1, \alpha_2 , \cdots) = k \text{     } \longrightarrow \text{     }\Lambda(\Delta) = \Lambda(\Delta,\alpha_0,\alpha_1,\alpha_2,\cdots)
\end{align*}

We could further derive and study the behaviors of these cost functionals over parameter space instead of reserve state. This would allow a preliminary framework to studying optimal parameterizations analytically, although given the complexity that arises from simple CPMM coupled markets, indicates this would likely require numerical methods.

While there are an immense amount of learnings from this model, understanding both the short and long-term springback effect will require expanding the model past a single swap.

\section{Conclusion}

While price coupling exists in traditional financial structures (this is essentially the definition of beta), the ability to do programmatic coupling at an asset level is novel. Issuance protocols such Doppler Protocol have enabled programmatic issuance. This new market structure enables a wider economic link through guaranteed liquidity and standardization via automated market makers. In short, both finding a liquid venue for an asset and running an auction both are incredible undertaking, thus the benefit of novel asset couplings has been outpaced by their cost.  Programmatic markets help lessen the burden on all parties through standardization and programming. 

Although there are added risks with coupling, we argue that a deeper understanding and adaptation of solutions from traditional financial markets will allow iterating to counteract the deleterious effects. With more analysis and more experimentation, market participants will slowly converge to better market structures and enable unique asset relationships that are totally novel.

\bibliography{references}

\begin{thebibliography}{1}

\bibitem{adams2025amammauctionmanagedautomatedmarket}
Austin Adams, Ciamac~C. Moallemi, Sara Reynolds, and Dan Robinson.
\newblock am-amm: An auction-managed automated market maker, 2025.

\bibitem{Angeris2021CFMM}
Guillermo Angeris, Akshay Agrawal, Alex Evans, Tarun Chitra, and Stephen Boyd.
\newblock Constant function market makers: Multi-asset trades via convex optimization.
\newblock {\em arXiv preprint}, 2021.

\bibitem{cieslak2021economics}
Anna Cieslak and Annette Vissing-Jorgensen.
\newblock The economics of the fed put.
\newblock {\em The Review of Financial Studies}, 34(9):4045--4089, 2021.

\bibitem{geanakoplos2010leverage}
John Geanakoplos.
\newblock The leverage cycle.
\newblock {\em NBER macroeconomics annual}, 24(1):1--66, 2010.

\bibitem{lehar2025decentralized}
Alfred Lehar and Christine Parlour.
\newblock Decentralized exchange: The uniswap automated market maker.
\newblock {\em The Journal of Finance}, 80(1):321--374, 2025.

\bibitem{milionis2025automatedmarketmakingarbitrage}
Jason Milionis, Ciamac~C. Moallemi, and Tim Roughgarden.
\newblock Automated market making and arbitrage profits in the presence of fees, 2025.

\bibitem{milionis2024automatedmarketmakinglossversusrebalancing}
Jason Milionis, Ciamac~C. Moallemi, Tim Roughgarden, and Anthony~Lee Zhang.
\newblock Automated market making and loss-versus-rebalancing, 2024.

\end{thebibliography}
 
\newpage
\section*{Notation}
\begin{align*}
    x, y, z && &\text{ Three assets forming the coupled market, where $y$/$x$ and $z$/$y$ are the trading pairs.} \\
    R = (x, y) && &\text{ Reserve state of an $y$/$x$ liquidity pool.} \\
    \phi(R) = k && &\text{ Constant function market maker invariant.} \\
    \Delta, \Lambda, \Gamma && &\text{ Amount of $x$, $y$, and $z$ assets involved in a swap context respectively, (input or output).} \\
    \gamma \in \left(0,1\right] && &\text{ Discount factor for the pool, $\gamma = 1 - \text{pool fee}$.} \\
    P_x(R), P_y(R) && &\text{ Spot price of the pool denominated in base asset $x$ or $y$.} \\
    \mu_y(\Delta) && &\text{ Relative price drift: $\mu_y = \frac{P'_y - P_y}{P_y}$.} \\
    D_{\text{marg},y}(\mu_y) && &\text{ Marginal swap depth, measuring liquidity absorption rate.} \\
    v(\mu_y) && &\text{ Value discrepancy between coupled and decoupled scenarios.}
\end{align*}

\end{document}